\documentclass[runningheads]{llncs}
\usepackage{graphicx}
\usepackage{amsmath}
\usepackage{amssymb} 
\usepackage{xspace}
\usepackage{array}
\usepackage[linesnumbered,noline,ruled,noend]{algorithm2e}
\usepackage{color}
\usepackage{xcolor}
\usepackage{listings}
\newcolumntype{L}{>{$}l<{$}}
\newcommand{\model}{\mathcal M}
\newcommand{\form}{\phi}
\newcommand{\props}{P}
\newcommand{\lneg}{\neg}

\newcommand{\lnear}{\mathcal{N}}

\newcommand{\msep}{\,\mid\,}
\newcommand{\slreach}[2]{\lreach \;#1[#2]}
\newcommand{\lreach}{\rho}

\usepackage{xspace}
\newcommand{\voxlogica}{{\tt VoxLogicA}\xspace}
\newcommand{\vlgpu}{{\tt VoxLogicA-GPU}\xspace}

\newcommand{\SLCS}{SLCS\xspace}
\newcommand\code[1]{\texttt{#1}}
\newif\ifLong
\Longtrue

\begin{document}
\title{A spatial model checker in GPU\ifLong\\(extended version)\fi
\thanks{Research partially supported by the MIUR Project PRIN 2017FTXR7S ``IT-
        MaTTerS''. 
        The authors are thankful to Raffaele Perego, Franco Maria Nardini and the 
        High Performance Computing Laboratory at ISTI-CNR for providing access 
        to the powerful machine used in our tests in Section \ref{sec:results}.   
        The authors also acknowledge fruitful discussions with Gina 
        Belmonte, Diego Latella, and Mieke Massink.
        }}
\author{ 
    Laura Bussi\inst{1} \and 
    Vincenzo Ciancia\inst{2}\and
    Fabio Gadducci  \inst{1}
}
\authorrunning{L. Bussi, V. Ciancia, F. Gadducci}

\institute{Dipartimento di Informatica, Universit\`a di Pisa\and
    Istituto di Scienza e Tecnologie dell'Informazione, CNR}

\maketitle

\begin{abstract}
    The tool \voxlogica merges the state-of-the-art library of computational 
    imaging algorithms
    \code{ITK} with the combination of declarative specification and
    optimised execution provided by spatial logic model checking.  The analysis of
    an existing benchmark for segmentation of brain tumours via a simple logical
    specification reached state-of-the-art accuracy. We present a new,
    GPU-based version of \voxlogica and discuss its implementation, scalability, 
    and applications.
\end{abstract}

\section{Introduction}

Spatial and Spatio-temporal model checking have gained an increasing
interest in recent years in various application domains, including collective
adaptive~\cite{Ciancia2018,CLMPV16} and networked
systems~\cite{Bartocci2016}, runtime monitoring~\cite{NBCLM15,BBLN17},
modelling of cyber-physical systems~\cite{TKG17} and medical imaging
images~\cite{Gr+09,Ba+20}. 
\ifLong 
Current research in this field has its origin in the
\emph{topological} approach to spatial logics, whose early theoretical foundations 
have been extended to encompass reasoning about \emph{discrete} spatial 
structures, such as
graphs and images \cite{CLLM16}, and reasoning over space-time
\cite{CLLM16bertinoro,TKG17,NBCLM15}, with related model checking algorithms and
tools \cite{CGLLM15,BCLM19,BartocciBLNS20}.
\fi

Introduced in~\cite{BCLM19}, 
\ifLong continuing the research line of
\cite{BCLM16,Ba+20,Be+17}, \fi 
 \voxlogica (\emph{Voxel-based
Logical Analyser})\footnote{\voxlogica: see
\url{https://github.com/vincenzoml/VoxLogicA}} caters  for a novel,declarative approach
to (medical) image segmentation, supported by spatial model checking. Therein, the main case
study was brain tumour segmentation for radiotherapy \ifLong(see
e.g.,~\cite{Fyllingen2016,Zhu2012,Despotovi2015,Simi2015,Dupont2016})\fi, 
using
the BraTS 2017 dataset
(\ifLong a publicly available set of benchmark MRI images
for brain tumour segmentation including high-quality \emph{ground truth}:
\fi
see~\cite{Bak+17}).
A simple high-level specification for glioblastoma segmentation was proposed and
tested using \voxlogica. The procedure competes in accuracy with
state-of-the-art techniques, most of which
based on machine learning.
This work presents a variant of \voxlogica, named \vlgpu, that implements
the core logical primitives in GPU. The motivation is shared with a recent trend
on theory and implementation of formal methods in GPU (see
\cite{BerkovichBF13,WijsB16,WijsNB16,NeeleWBP16,OsamaW19b,OsamaW19}): 
to take advantage of the availability of high-performance, massively parallel
computational devices, 
in order to 
increase the size of tractable problems. We describe the tool implementation,
architecture, and issues (in particular, connected component labelling in GPU).
We compared the CPU and GPU implementations, both on artificial test cases
consisting of very large formulas, and on the brain tumour segmentation case
study. Results aimed at checking scalability on large formulas are very
encouraging, as the command pipeline of the GPU is fully exploited, and there is
a considerable speed-up with respect to using  the CPU. On the case study, the
current limitations of \vlgpu (namely, the restriction to 2D images and the
smaller number of primitive operations that has been implemented) forbid a
direct comparison with \cite{BCLM19}, but the GPU version was still able to
outperform the CPU on a simplified experiment, which is particularly relevant,
as the CPU algorithms rely on a state-of-the-art imaging library.

\ifLong
\section{The Spatial Logic \SLCS}
\label{sec:slcs}

In this section, we briefly review the syntax of the spatial logic \SLCS,
defined in \cite{CLLM14,CLLM16}, and its interpretation, restricted to the case
of two-dimensional images which is currently handled by VoxLogicA-GPU. For the
general definition on so-called \emph{closure spaces}, and the link between
(multi-dimensional) images, graphs, closure spaces and topological spatial
logics we refer the reader to \cite{CLLM16,Ba+20,BCLM19}. The syntax of the
logic we use in this paper is its most up-to-date rendition, where the
\emph{surrounded} connective from \cite{CLLM16} is a derived one, whereas
reachability is primitive, as in \cite{BBLN17,CLM19,CLMV20}. Given set $\props$
of {\em atomic predicates}, with $p \in \props$, the syntax of the logic is
described by the following grammar:
\begin{equation}\label{def:syntax}
    \form  ::=   p  \msep  \lneg \, \form \msep  \form_1 \, \land \, \form_2  \msep  \lnear \form \msep \slreach{\form_1}{\form_2}
\end{equation}

The logic is interpreted on the pixels of an image $\model$ of fixed dimensions.
The truth values of a formula $\form$ on \emph{all} the pixels can be rendered
as a binary image of the same dimensions of $\model$. Therefore, in particular,
\textbf{atomic propositions} correspond to binary images. To get an intuition,
consider that typical atomic propositions are numeric constraints (thresholds)
on imaging features, such as intensity, or red, green, blue colour components.
\textbf{Boolean operators} are defined pixel-wise: $\lnot \form$ is the
complement of the binary image representing $\form$, and $\form_1 \land \form_2$
is binary image intersection. The \textbf{modal} formula $\lnear \form$ is
interpreted as the set of pixels that share a vertex or an edge with any of the
pixels that satisfy $\form$ (adopting the so-called \emph{Moore neighbourhood});
in imaging terminology, this is the \emph{dilation} of the binary image
corresponding to $\form$. The \textbf{reachability} formula
$\slreach{\form_1}{\form_2}$ is interpreted as the set of pixels $x$ such that
there is a pixel $y$, a path $\pi$ and an index $\ell$ such that $\pi_0 = x$,
$\pi_\ell = y$, $y$ satisfies $\form_1$, and all the intermediate pixels $\pi_1,
\ldots, \pi_{\ell - 1}$ (if any) satisfy $\form_2$.

From the basic operators, several interesting notions can be derived, such as
\emph{interior} (corresponding to the imaging primitive of \emph{erosion}),
\emph{surroundedness}, \emph{contact} between regions (see also \cite{CLM19},
encoding the \emph{discrete} Region Calculus RCC8D of \cite{RLG13} in a variant
of SLCS).
\fi
\section{The tool \vlgpu}
\label{sec:implementation}

\vlgpu is a \emph{global}, \emph{explicit state} model checker for the logic SLCS defined in \cite{CLLM16} , aiming at high efficiency and maximum portability
\vlgpu is implemented in \texttt{FSharp},
using the \texttt{NET Core} infrastructure, and
the \emph{General-Purpose GPU computing} library \texttt{OpenCL}.\footnote{\texttt{FSharp}: see \url{https://fsharp.org}. \texttt{NET Core}:
see \url{https://dotnet.microsoft.com}. \texttt{OpenCL}: see
\url{https://www.khronos.org/opencl}. \texttt{ITK}: see \url{https://itk.org}.} Efficiency is
one of the major challenges, as outperforming \voxlogica inherently means
designing \emph{in-GPU} imaging primitives faster than the state-of-the-art
library ITK. 
The focus of the first release of \vlgpu is on the implementation and early
dissemination of a free and open source infrastructure to experiment with
GPU-based spatial model checking, and demonstrate its scalability. Thus,
development has been narrowed to a core implementation that is powerful enough
to reach the stated objectives, although not as feature-complete as \voxlogica.
IN particular, the implemented primitives are those of \SLCS, plus basic
arithmetic. More complex operations (normalization, distance transform,
statistical texture analysis) are yet to be implemented. Furthermore, in the
first release, computation is restricted to 2D images and 16 bit unsigned
integers\footnote{The precision of 16 bits is needed for the BraTS dataset, see
Section \ref{sec:results}.}; this eased development, as a
separate kernel is needed for each dimensionality and each numeric type.

\ifLong{\subsection{Syntax}}\fi

\vlgpu is a command-line tool. It takes only one parameter, a text file
(usually, with \texttt{.imgql} extension) containing the specification to be
executed.  
\ifLong 
In the following, \code{f, x1,\ldots, xN, x} are identifiers,
\code{"s"} is a string, and \code{e1, \ldots, eN, e} are expressions (to be
detailed later). 
\fi 
A specification consists of a text file containing a sequence
of \textbf{commands} (see Section~\ref{sec:results-brats} for an example). Five
commands are currently implemented: \texttt{let}, \texttt{load}, \texttt{save},
\texttt{print}, \texttt{import}. 
\ifLong
For the scope of this work, it suffices to
describe the first three. The command \code{let f(x1,...,xN) = e} is
\emph{function declaration} (\code{let f = e} declares a constant) with a
special syntax for \emph{infix} operators; the command \code{load x = "s"} loads
an image from file \code{"s"} and binds it to \code{x} for subsequent usage; the
command \code{save "s" e} stores the image resulting from evaluation of
expression \code{e} to file \code{"s"}.
\vlgpu comes equipped  with built-in arithmetic
operators and logic primitives.
Furthermore, a ``standard library'' is provided containing short-hands for
commonly used functions, and for derived operators. 
An \textbf{expression} may be a number (with no distinction between
floating point and integer constants), an identifier (e.g. \code{x}), a function
application (e.g. \code{f(x1,x2)}), an infix operator application (e.g. \code{x1
+ x2}), or a parenthesized expression (e.g. \code{(x1 + x2)}). 
\fi

\ifLong\subsection{Implementation details}\fi

The core model-checking algorithm of \vlgpu is in common with \voxlogica. After
parsing, \ifLong imported libraries are resolved, load instructions are
executed, and \fi parametric macros are expanded, avoiding to duplicate
sub-expressions (see \cite{BCLM19} for more details). The result is a
\emph{directed acyclic graph}, where each node contains a task to execute, after
the tasks from which it depends (denoted by edges) are completed. A task can be
either an operator of the language, or an output instruction. The semantics of
operators is delegated to the specific implementation of the VoxLogicA API,
which must define the type \texttt{Value}, on which tasks operate.
Operators are implemented in a module running on CPU that launches one or more
\emph{kernels} (i.e. functions running on GPU), executed in an asynchronous way. 
As the execution order is not known a priori, the desired kernel is retrieved at runtime 
and passed to an auxiliary module that performs the actual call to the GPU code. 
This module is also in charge of setting the actual parameters (i.e. image buffers or
numeric values) of the kernel.
The type Value is used to describe information about the type of data stored in
the GPU memory (a binary or numeric image, or a number). More precisely, it
contains a pointer to an OpenCL buffer stored in the GPU memory, the pixel type
of the image and the number of components per pixel. These information are
needed in order to have fully composable operators, as the image type may vary
during computation (e.g. when computing connected components labelling).
Providing a pointer to the OpenCL buffers minimises data transferring between
host and device, as data is retrieved from the CPU only when it is strictly
necessary to continue execution.
The correct execution order is preserved using OpenCL \emph{events}, i.e.
objects that communicate the status of a command. Due to asynchronicity, kernels
are pushed in a \emph{command queue} that may also contain global barriers or
read/write to the GPU memory. Each time a command is added to the queue, a new
event is added to the event list and will be eventually fired at the command
completion. Events can be chained, and the CPU can wait for an event to complete
and read the result of an associated computation from the GPU. Wait instructions
are only used in the implementation of reachability-based logical primitives, in
order to check if the GPU reported termination of an iterative algorithm, and
when saving results, achieving very high GPU usage.
Just like in $\voxlogica$, the reachability operator $\slreach{\form_1}{\form_2}$ is
implemented using connected components labelling, which makes the implementation such operation in GPU very relevant for this paper.\ifLong In order to get an intuition
about this, consider a binary image $I$ whose non-zero points are the points
satisfying $\form_2$. Consider any pixel $z$ in a connected component of $I$
that has a point in common with the dilation of $\form_1$. By  construction
there is a path $\pi$ and an index $\ell$, with $\pi_0 = z$, $\pi_\ell$
satisfying $\form_1$, and all the pixels $\pi_0, \ldots, \pi_{\ell - 1}$
satisfying $\form_2$. Therefore the points in the dilation of the set of all $z$
in such situation satisfies $\slreach {\form_1}{\form_2}$, as well as the pixels
satisfying $\lnear \form_1$; there are no other pixels satisfying
$\slreach{\form_1}{\form_2}$.\fi

\subsection{Connected components labelling in \vlgpu}
The current, iterative algorithm for connected component labelling is
based on the \emph{pointer jumping} technique, and it has been 
designed to balance implementation simplicity with
performance. A more detailed investigation of this and other algorithms is left for future
research.
Algorithm~\ref{alg:CCL} presents the pseudo-code of the kernels 
(termination checking is omitted); see Figure~\ref{fig:cc-movie} for an example. 

\enlargethispage*{\baselineskip}
\enlargethispage*{\baselineskip}
\begin{algorithm} 
    \DontPrintSemicolon
    \textbf{initialization}(start: image of bool, output: image of int $\times$ int )\\
        \Indp{\For{$(i,j) \in Coords$}{
            \uIf{start(i,j)}{
                $output(i,j) = (i,j)$ // null otherwise
            }
        }} \Indm

    \textbf{mainIteration}(start: image of bool, input, output: image of int $\times$ int)\\
        \Indp{\For{$(i,j) \in Coords$}{
            \uIf{start(i,j)}{
                $(i',j') = input(i,j)$\;
                $output(i,j) = maxNeighbour(input,i',j')$
            }
        }} \Indm

    \textbf{reconnect}(start: image of bool, input, output: image of int $\times$ int)\\
    \Indp{\For{$(i,j) \in Coords$}{
        \uIf{start(i,j)}{
            $(i',j') = input(i,j)$\;
            $(a,b) = maxNeighbour(input,i,j)$\;
            $(c,d) = input(i',j')$\;
            \uIf{$(a,b) > (c,d)$}{
                $output(i',j') = (a,b)$ // Requires atomic write
            }
        }
    }}
    \caption{\label{alg:CCL}Pseudocode for connected components labelling}
\end{algorithm}

\ifLong
The algorithm uses coordinates
$(x,y)$ as labels. Starting from a binary image $I$, the algorithm initialises
the output image (a two-dimensional array of pairs) by labelling each non-zero
pixel of $I$ with its coordinates. At each iteration, in parallel on the GPU,
for each $p$ at $(x,y$) which is non-zero in $I$, containing $(x',y')$, the
Moore neighbourhood of $(x',y')$ is inspected to find a maximum value
$(x'',y'')$; the lexicographic order is used (possibly, $(x'',y'') = (x',y')$).
The value $(x'',y'')$ is written at $(x,y)$. 
See Figure~\ref{fig:cc-movie} for an example. 

Such algorithm converges in a logarithmic number of iterations with respect to
the number of pixels due to pointer jumping. The algorithm may fail to label
uniquely those connected components that contain a particular type of ``concave
corner'', namely a pixel at coordinates $(a,b)$, together with the pixels at
$(a+1,b)$, $(a,b+1)$, but not the pixel at $(a+1,b+1)$. 


To overcome this, each $k$ main iterations (with $k$ very small, $k=16$ in the current
implementation) the algorithm employs a \emph{reconnect} step, that inspects the
neighbourhood of each pixel $p$, containing $(x,y)$, looking for a pair
$(x',y')$ greater than the pair stored at $(x,y)$. In that case, $(x',y')$ is
written at $(x,y)$, and the main iterations are restarted, which immediately
propagates $(x',y')$ to $p$, and all other pixels that contain $(x',y')$.  At
this point termination is checked, by invoking a separate kernel that checks if
any pixel which is true in the original binary image has a neighbour with a
different label (if no pixel is in this situation, then the connected components
have been properly labelled)

By construction,
the algorithm exits \emph{only if} all connected components have been correctly
labelled.  An invariant is that after each main iteration, each pixel $p$
containing $(x,y)$ lays in the same connected component as $(x,y)$. For
termination, call $k$ the number of distinct labels in the image along
computation. Clearly, $k$ does not increase after a main iteration (in most
iterations it decreases). After each reconnect step, there are three
possibilities: 1) $k$ decreases; 2) $k$ does not change, but in the next step,
the algorithm terminates; 3) $k$ does not change, but in the next main
iteration, $k$ decreases. Therefore, $k$ always decreases after a reconnect step
and a main iteration, unless the algorithm terminates. Thus, by the invariant,
$k$ converges to a minimum which is the number of connected components.
In practice, convergence is quite fast (usually, logarithmic in
the number of pixels) and suffices to
establish the results in Section~\ref{sec:results}.
\fi

\ifLong
\begin{figure}
    \centering
    \begin{tabular}{cccc}
        \includegraphics[width=.25\textwidth]{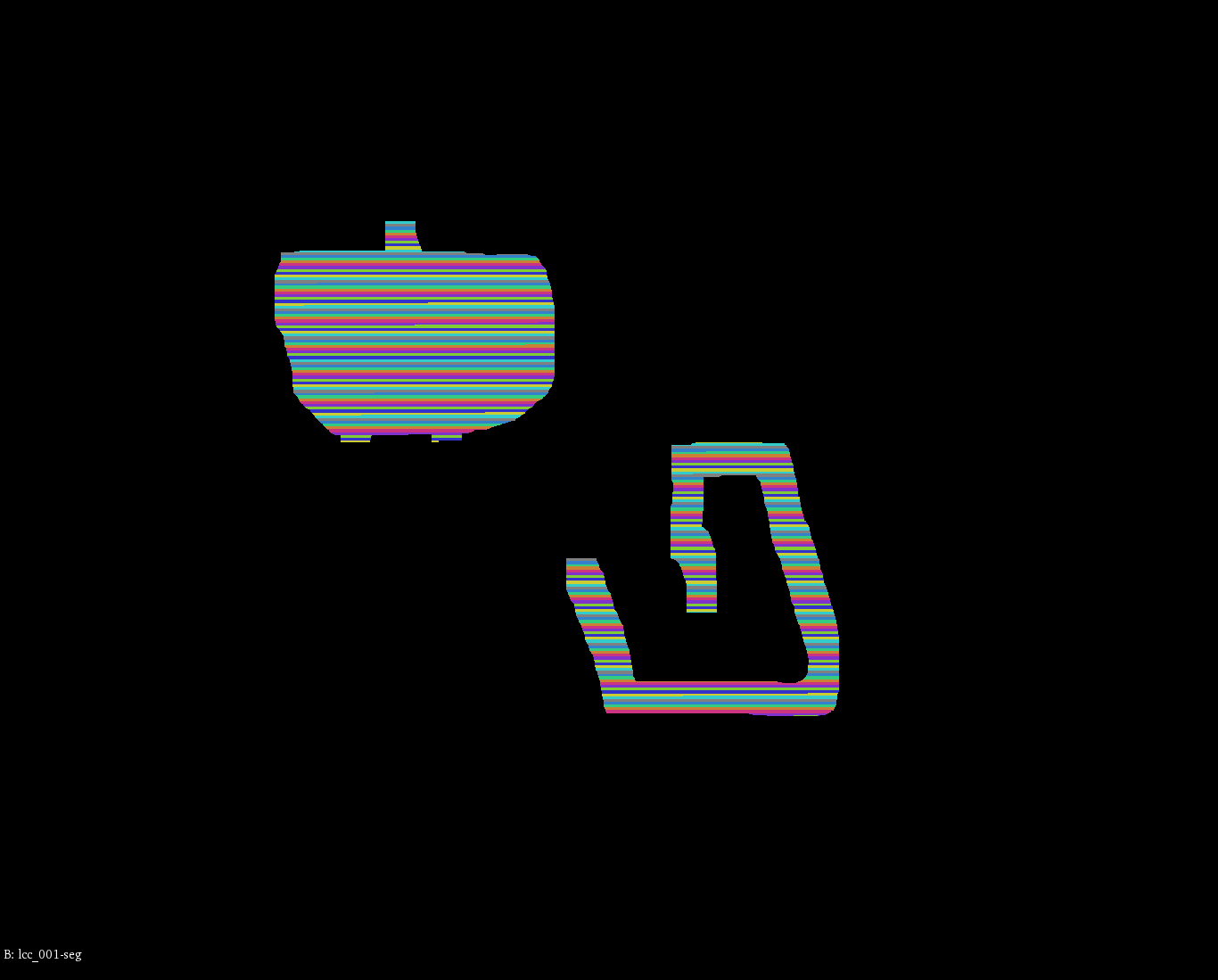} & 
        \includegraphics[width=.25\textwidth]{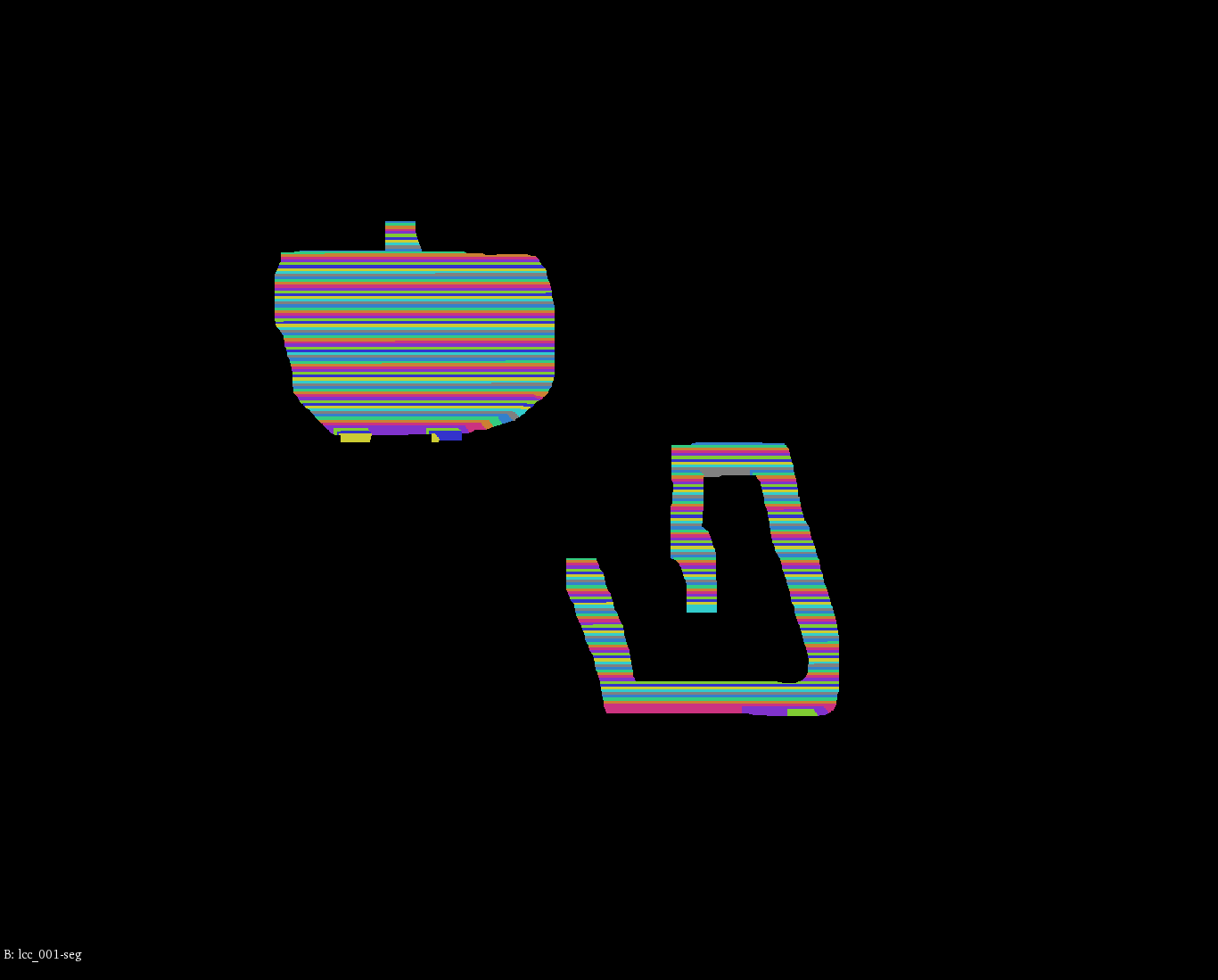} & 
        \includegraphics[width=.25\textwidth]{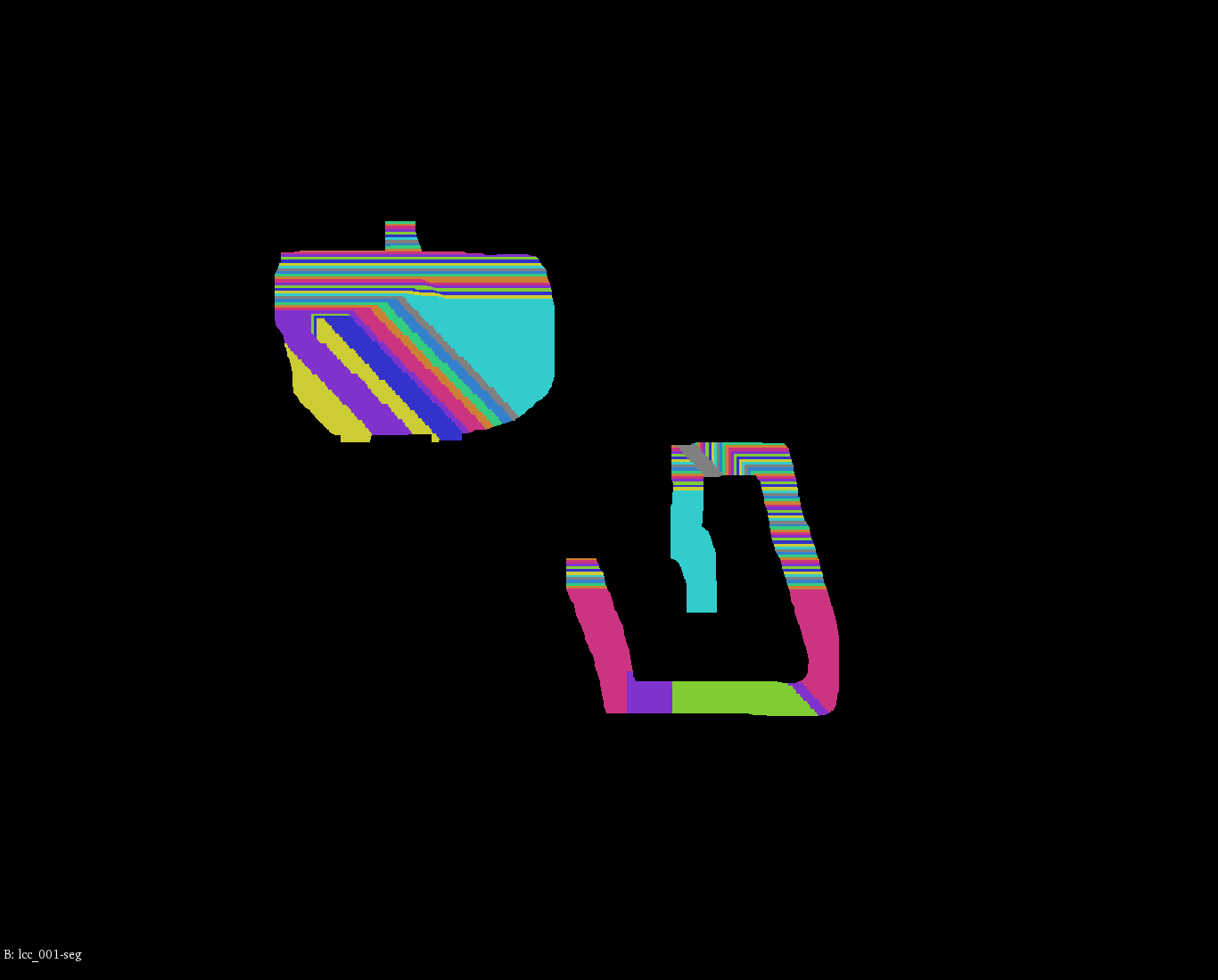} &
        \includegraphics[width=.25\textwidth]{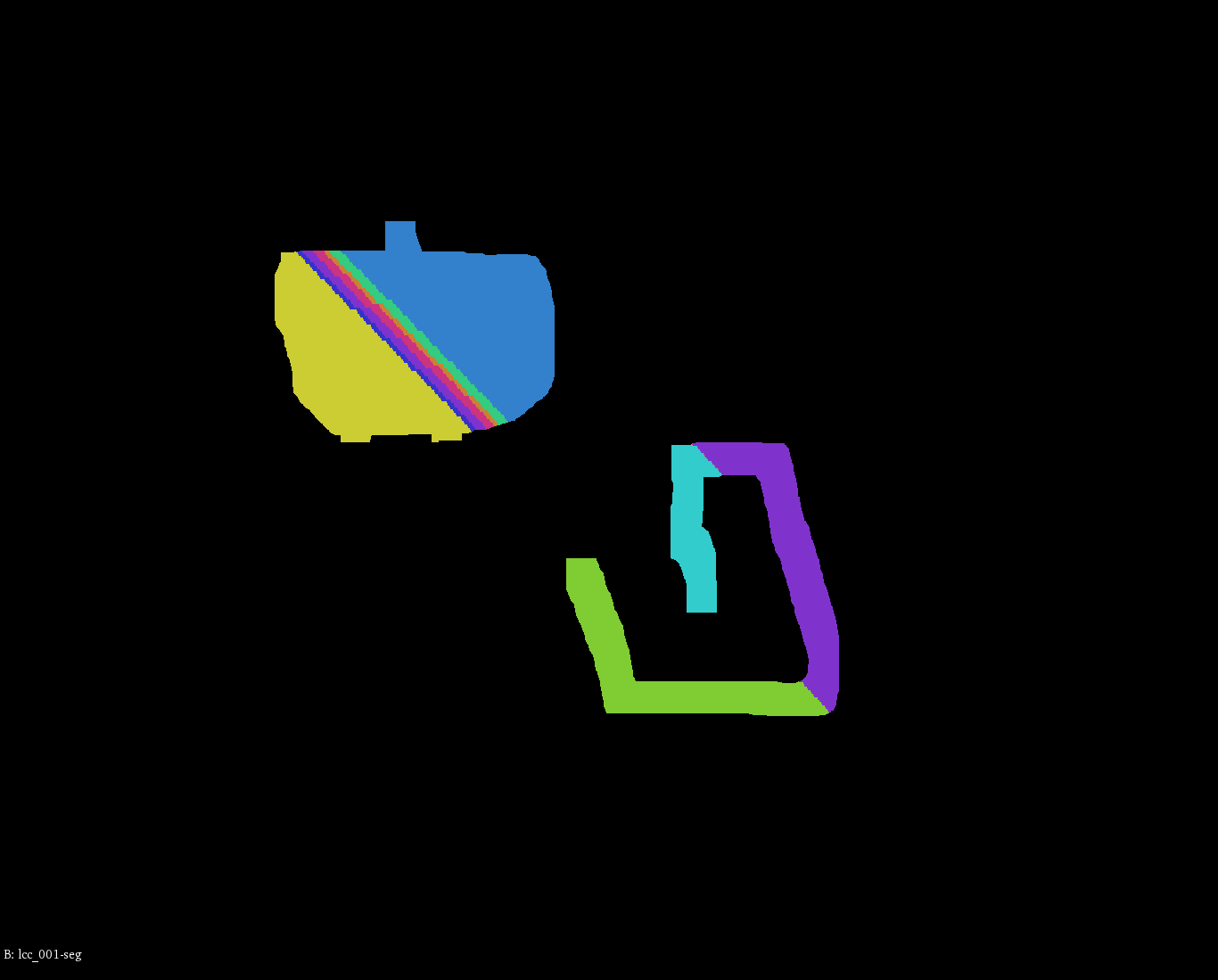} \\        
        Iteration 1 & Iteration 5 & Iteration 9 & Iteration 13 \\ 
        \includegraphics[width=.25\textwidth]{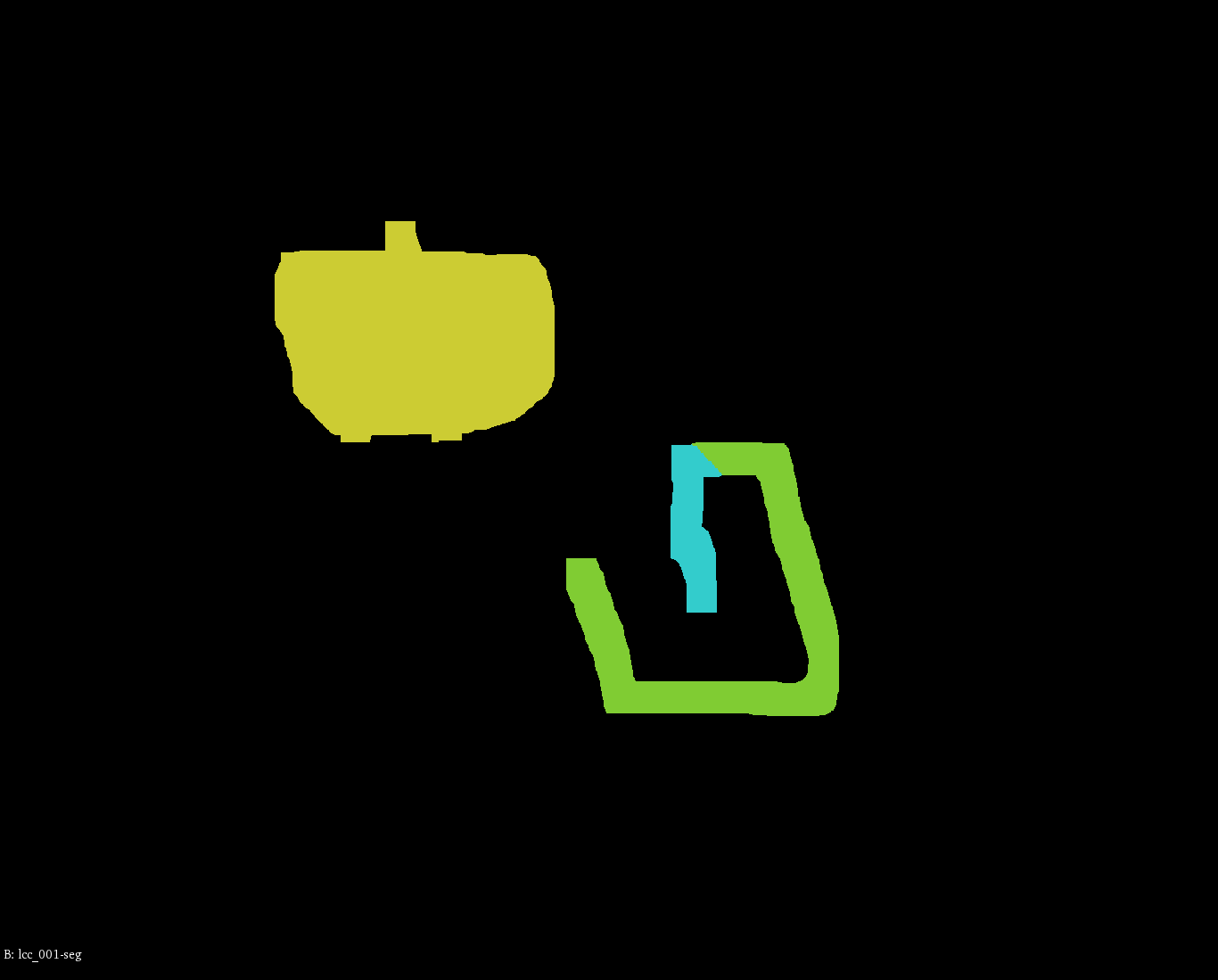} & 
        \includegraphics[width=.25\textwidth]{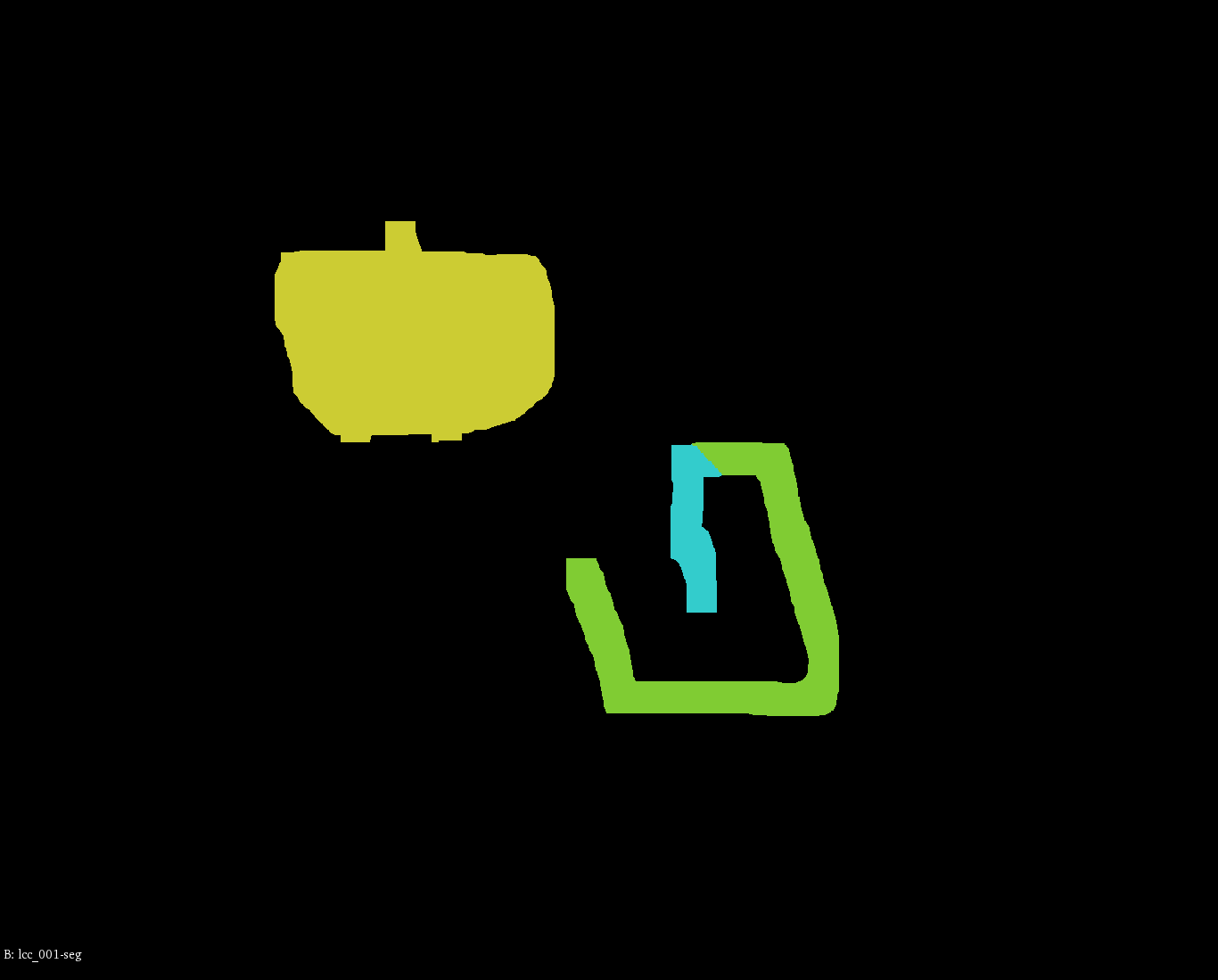} & 
        \includegraphics[width=.25\textwidth]{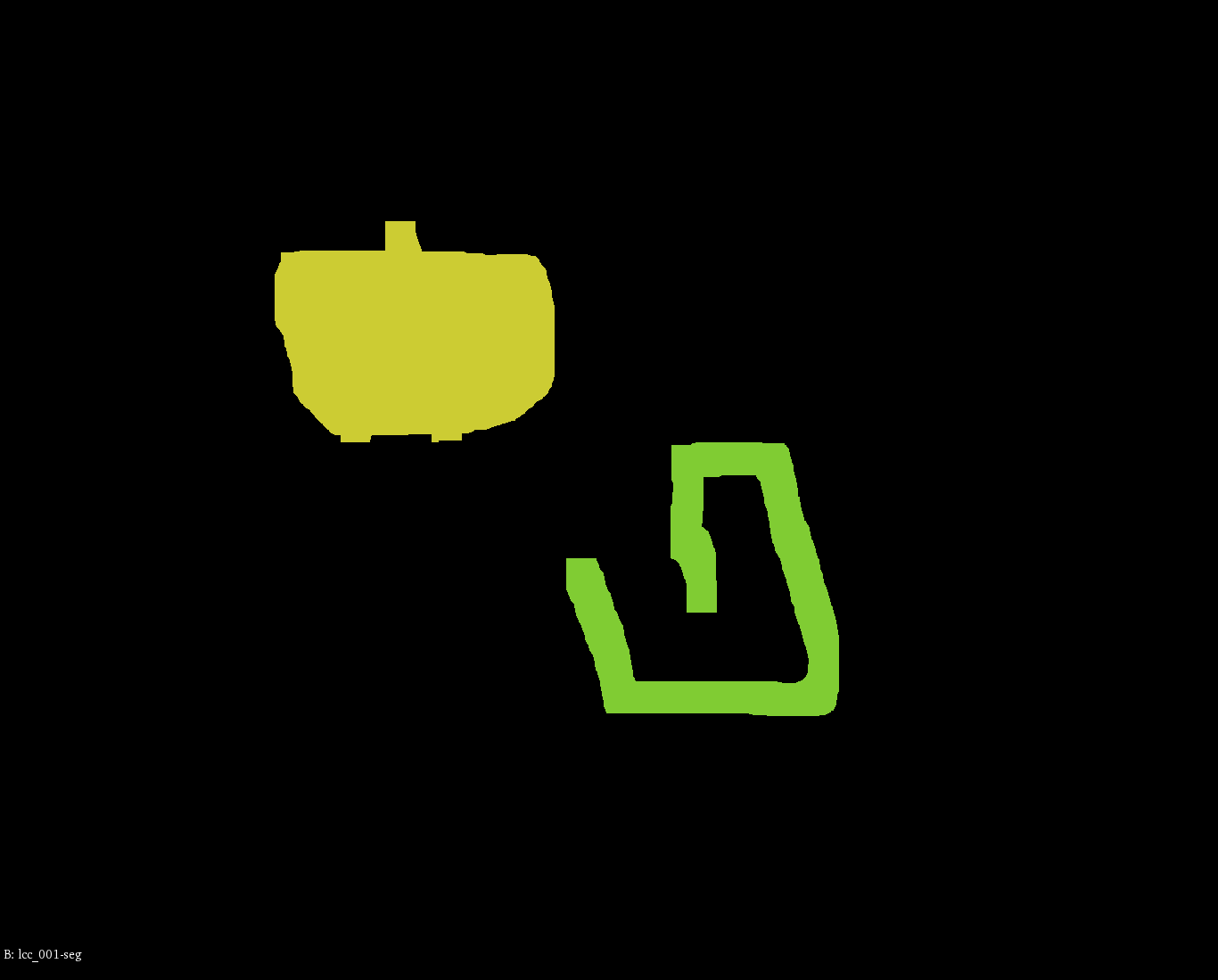} &
        \includegraphics[width=.25\textwidth]{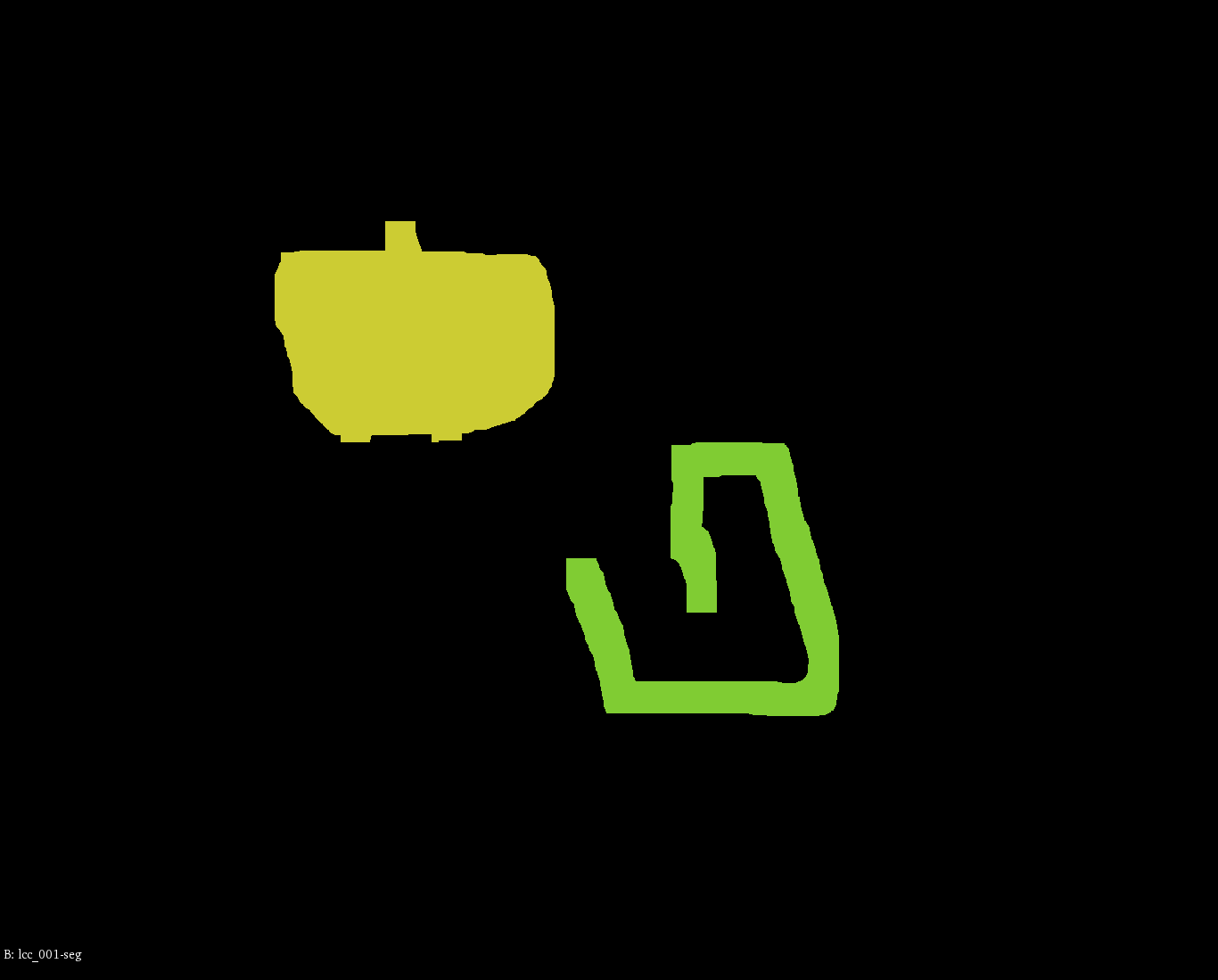} \\        
        Iteration 17 & Iteration 21 & Iteration 25 & Iteration 29
    \end{tabular}

    \caption{\label{fig:cc-movie}CC-labelling algorithm on a test image. Different colours represent different labels. Reconnect is called every 8 main iterations. At iteration 13, the main iterations have converged, therefore the image stays unchanged until iteration 16 (reconnect); iteration 17 shows label propagation after reconnect. Then until iteration 24 (last reconnect), images do not change again.}
\end{figure}
\fi

\section{Tests and Brain Tumour Segmentation Case Study}
\label{sec:results}

\subsection{Performance on large formulas}
\label{sec:results-large-formulas}

\begin{figure}\label{img:perfseq}    
        \includegraphics[scale=0.27]{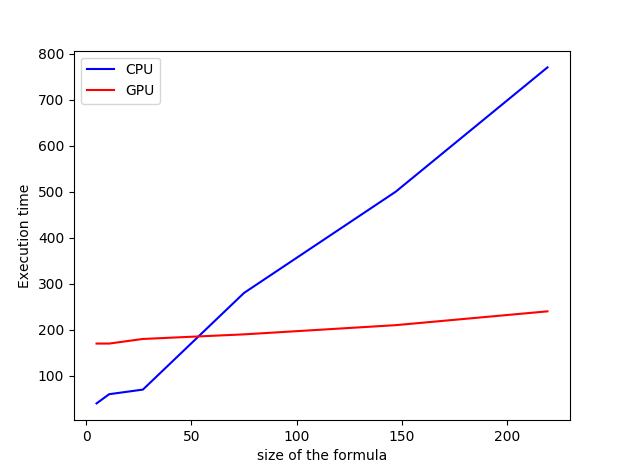}
        \includegraphics[scale=0.27]{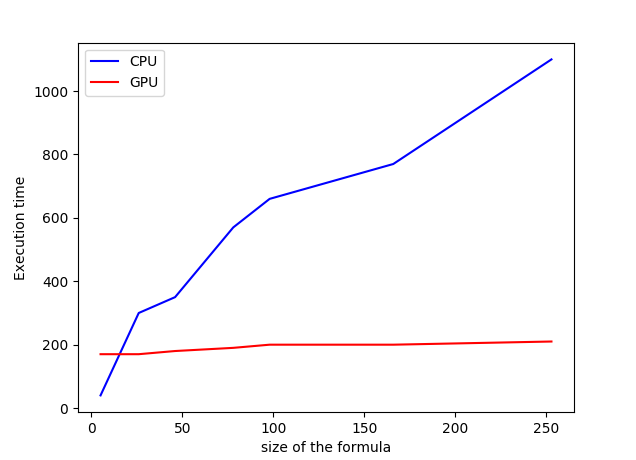}
    \caption{Performance on growing sequential (left) and random (right) formulas}
\end{figure}

We built two kinds of large formulas in order to stress the infrastructure:
sequential forumlas (i.e. formulas of the form f(g(...(x)))), and ``random''
ones, were all the operators are composed in various ways and applied over
different images. In both these cases, \vlgpu scales better than the CPU version
by a linear factor as the size of the formula grows. It is worth noting that the
CPU version has better performances on very small formulas in the random test,
due to the overhead needed to set up GPU computation. Figure~\ref{img:perfseq}
shows how the CPU and the GPU versions scale differently on growing formulas. As
in Section~\ref{sec:results-brats}, execution times are small. We foresee more
important speed-ups on real-world experiments, in the near future.
\subsection{Brain Tumour Segmentation}
\label{sec:results-brats}

The main case study of \cite{BCLM19} was brain tumour segmentation for
radiotherapy, achieving accuracy results on par with the state of the art
tools.
Validation was performed on the BraTS 2017
dataset\footnote{See \url{http://braintumorsegmentation.org}}, which provides
\emph{MRI-flair} 3D images composed by circa 9 millions of \emph{voxels}
(three-dimensional pixels) and manual segmentations for accuracy
scores.
Given the current restrictions of \vlgpu to 2D images, core logical primitives,
and 16-bit unsigned integers (see Section \ref{sec:implementation}), 
we have extrapolated a simplified dataset 
and the specification in Figure~\ref{fig:specification}.
\begin{figure}
   \centering
 \begin{lstlisting}
// Load data (16 bit png, normalised)
load img = "normalised-MRI-flair.png" 

// 1. Thresholds
let hI = intensity(img) >. 62258 // (62258 = 0.95 * 65535; hyperintense)
let vI = intensity(img) >. 56360 // (56360 = 0.86 * 65535; very intense)

// 2. Semantic noise removal via region growing
let gtv = grow(hI,vI)

// Save the results
save "segmentation.png" gtv
\end{lstlisting}
   \caption{\label{fig:specification} Brain Tumour Segmentation in \vlgpu}
\end{figure}

The specification uses the operator
$grow(\form_1,\form_2)$, derived from $\rho$, to identify pixels that either
satisfy $\form_1$, or belong to a region which is in contact with $\form_1$, and
only contains pixels satisfying $\form_2$. The particular simplicity of such
procedure should not raise doubts about its effectiveness, as ``semantic noise
removal'' is the core of the specification in \cite{BCLM19}, and contributes to
its accuracy (and computational cost) for the main part; but note that the
restriction to 2D images makes the procedure too fast to be measured in smaller
images, requiring artificial up-scaling in order to compare the performance of
\voxlogica and \vlgpu.

\ifLong
To extract a 2D slice out of each 3D MRI-flair scan of the BraTS dataset, each 3D 
image is first normalised using
the \texttt{percentiles} operator of VoxLogicA, in order to remain faithful to
the kind of processing done in \cite{BCLM19}. The normalised image has floating
point intensity values between 0 and 1; these have been multiplied by $2^{16}-1$
(the maximum 16-bit unsigned integer) in order to maximise precision (therefore
the thresholds in the specification were also multiplied by the same number).
Finally, the ``most significant'' slice has been selected as the one where the
manual segmentation has a larger volume, resulting in an image which contains
enough information to test the performance of \vlgpu. 
Finally, the slice has been up-scaled (with no filtering) in order to obtain a
very large image, (size is $7680\times 7680$ pixels, that is circa $60$
megapixels) in order to challenge the efficiency of VoxLogicA, where the
implementation of most logical primitives is delegated to the state-of-the-art
imaging library ITK\footnote{The Insight Toolkit, see \url{https://itk.org/}.}.
\fi

We have tested the specification on 4 devices, averaging the measurements on
several runs.  The GPU algorithm has been tested with \vlgpu on two different
GPUs, namely an \emph{Intel HD Graphics 630} (integrated GPU with very low
performance), taking about 3 seconds to complete;  and a \emph{NVIDIA TITAN Xp
GP102}, which is a high-end desktop GPU, taking about 600ms to complete. The CPU
algorithm has been tested with \voxlogica on the same machine used in
\cite{BCLM19}, equipped with a \emph{Intel Core i7-7700} CPU, taking about 1100
milliseconds, and on the machine hosting the \emph{NVIDIA TITAN Xp}, which also
has a \emph{Intel Xeon E5-2630} CPU, taking 750 milliseconds. 
\ifLong
Runtime was
measured from the moment in which the tools print ``starting computation'' in
the log (signifying that all preparation, parsing, memoization, etc. has been
finalised) to the moment in which the tools print ``saving file ...'', meaning
that the results have been fully computed, and transferred from the GPU to the
CPU (for \vlgpu).
\fi
\ifLong
The measured times are quite low, hence not very significant, due to
the simplicity of the procedure and the restriction to 2D images. Doubling the
image size is not possible at the moment as it would not be possible to allocate
the needed memory buffers on the devices available at the moment.
Obviously, larger and more complex specifications, such as the gray and white
matter segmentation in~\cite{BelmonteCLM19}, which takes minutes to complete
in CPU, will benefit more of the GPU implementation (as discussed in
Section~\ref{sec:results-large-formulas}). More relevant tests will be possible
when the tool will support 3D images (which are more challenging for the CPU)
and after the implementation of further, heavy operations of \voxlogica such as
local histogram analysis. It is noteworthy that the GPU implementation already
outperforms the CPU in this case study (on the available devices), and that both
implementations achieve impressive practical results, by segmenting a brain
tumour on such a large image in less than one second. In CPU, this is due to
the state-of-the art implementation of primitives in the ITK library. In GPU,
the connected components algorithm that we designed yields very good results.
\fi

\section{Conclusions}

Our preliminary evaluation of spatial model checking in GPU is encouraging:
large formulas benefit most of the GPU, with significant speed-ups. Connected
components labelling should be a focus for future work; indeed, the topic is
very active, and our simple algorithm, that we used as a proof-of-concept, may
as well be entirely replaced by state-of-the-art, more complex procedures (see
e.g. the recent work~\cite{AllegrettiBG20}). 
\ Making \vlgpu feature-complete with
respect to \voxlogica is also a stated goal for future development. In this respect,
we remark that although in this work we decided to go through the ``GPU-only''
route, future development will also consider a \emph{hybrid} execution mode with
some operations executed on the CPU, so that existing primitives in \voxlogica
can be run in parallel with those that already have a GPU implementation.
Usability of \vlgpu would be greatly enhanced by an user interface; however,
understanding modal logical  formulas is generally considered a difficult task, and
\ifLong
 the novel combination of declarative programming and domain-specific medical
image analysis concepts may hinder applicability of the methodology, as
\fi
cognitive/human aspects may become predominant with respect to technological
concerns. In this respect, we plan to investigate the application of formal
methodologies (see e.g. \cite{BrocciaMO19}).

\bibliographystyle{splncs04}
\bibliography{bibliography}

\end{document}